\documentclass[a4paper,reprint,prd,floatfix,superscriptaddress]{revtex4-2}
\usepackage[T1]{fontenc}
\usepackage[utf8]{inputenc}
\usepackage{lmodern}
\usepackage{graphicx}
\usepackage{dcolumn}
\usepackage{bm}
\usepackage{bbm}
\usepackage{mathrsfs}
\usepackage{eufrak}
\usepackage{amsmath}
\usepackage{amsfonts}
\usepackage[normalem]{ulem}
\usepackage[colorlinks=true,linkcolor=blue,citecolor=blue,urlcolor=blue]{hyperref}

\usepackage{amssymb}
\usepackage{subfigure}
\usepackage{url}
\usepackage{color}
\usepackage[ddmmyy,24hr]{datetime}
\usepackage{bigdelim}
\usepackage{booktabs}
\usepackage{multirow}
\usepackage{subfigure}
\usepackage{physics}
\usepackage{cancel}
\usepackage{stackrel}
\usepackage{paralist}
\usepackage{xspace}
\usepackage{slashed}
\usepackage{enumerate}
\usepackage{float}
\usepackage{fullpage}
\usepackage{wasysym}
\usepackage{comment}
\usepackage{bbold}
\usepackage{orcidlink}

\begin{document}

\title{The effect of nuclear recoil on neutrino oscillations: Toward understanding of short baseline anomalies}

\author{L. Parini \orcidlink{0000-0000-0000-0000}}
\email{[luca.parini@unisalento.it]}
\affiliation{Dipartimento di Matematica e Fisica, Università del Salento and INFN Sezione di Lecce}
\affiliation{Institute of Nanotechnology of the National Research Council of Italy, CNR-NANOTEC, Lecce Central Unit, c/o Campus Ecotekne, Via Monteroni, 73100 Lecce, Italy}

\author{E. V. Stefanovich \orcidlink{0000-0002-8973-7804}}
\email{[eugene_stefanovich@usa.net]}
\noaffiliation

\begin{abstract}
We studied the structure of the neutrino wave functions produced in nuclear decays, with particular emphasis on the role of nuclear recoil. Although the fraction of the recoil energy associated with a nonzero neutrino mass is extremely small, it gives rise to a notable time-dependent flavor oscillation. For long-lived sources, such as those used in gallium anomaly and reactor anomaly experiments, this recoil-driven oscillation makes a substantial contribution to the observed deficit of (anti)neutrinos.
\end{abstract}

\maketitle

\section{Introduction}
\label{sec:introduction}

Neutrino oscillations have been detected in solar, atmospheric, reactor, and accelerator experiments. In the majority of cases, the observations are well explained by the standard three-flavor framework. Nonetheless, current theoretical models face difficulties in accounting for several short-baseline results \cite{Acero-2024}, with the gallium anomaly \cite{SAGE, GALLEX, BEST, Gallium_anomaly} standing out as particularly puzzling. 

As an illustration, in the BEST experiment \cite{BEST} neutrinos were generated via electron-capture decays of chromium-51,

\begin{align}
^{51}\!\mathrm{Cr} &\to ^{51}\!\mathrm{V}+\nu_e. \label{eq:Cr}
\end{align}

\noindent By lepton number conservation, this process can yield only electron neutrinos. Neutrino oscillations, of course, imply that lepton number conservation is not an exact symmetry of nature. Nonetheless, it is commonly assumed that flavor conversion becomes significant only after the neutrino has traveled a distance comparable to its oscillation length. The intriguing aspect of short-baseline anomalies is that they point to flavor conversion over much shorter path lengths. For instance, the oscillation length for neutrinos produced in reaction (\ref{eq:Cr}) lies between 740 m and 25 km, yet experiments observe that roughly 20\% of the $\nu_e$ flux vanishes over distances shorter than 2 meters.

A frequently proposed way to account for short-baseline anomalies is to introduce an additional sterile neutrino species with a larger mass and correspondingly shorter oscillation length. Yet, detailed studies have shown that a single sterile neutrino cannot consistently account for all available data, implying that more complex frameworks are necessary \cite{Banks, Brdar, Farzan-2023, Hardin, Giunti-2022, Giunti-2024, Acero-2024, Acharya}.

In this work, we aim to address short-baseline anomalies without postulating new particles or relying on unverified physical mechanisms. Our focus is on the process of neutrino production in decays, with particular attention to nuclear recoil. Although the portion of the recoil energy arising from the nonzero neutrino mass is extremely small, it induces previously neglected temporal oscillations in flavor probabilities, which may account for short-baseline anomalies.

In Section~\ref{sc:plane}, we introduce a simplified plane-wave framework to illustrate how the internal composition of the neutrino state influences its oscillation pattern. Section~\ref{sc:Application} then uses these findings to address short-baseline anomalies.

Throughout, we adopt natural units with $\hbar = 1$ and $c = 1$, except where they are written explicitly. For clarity, our analysis is restricted to a single spatial dimension.

\section{Plane wave solutions}
\label{sc:plane}

Wave functions of neutrinos generated in two-body decays exhibit an exponential profile that moves away from the emission point with (nearly) the speed of light \cite{Pavlichenkov} 

\begin{align*}
\Psi(r, t) &= \mathcal{N} e^{-i \mathcal{E} t }e^{-\Gamma (t- r)/2} \theta(t-r). 
\end{align*}

\begin{figure} [h!]
\includegraphics[width=8cm,height=4.5cm] {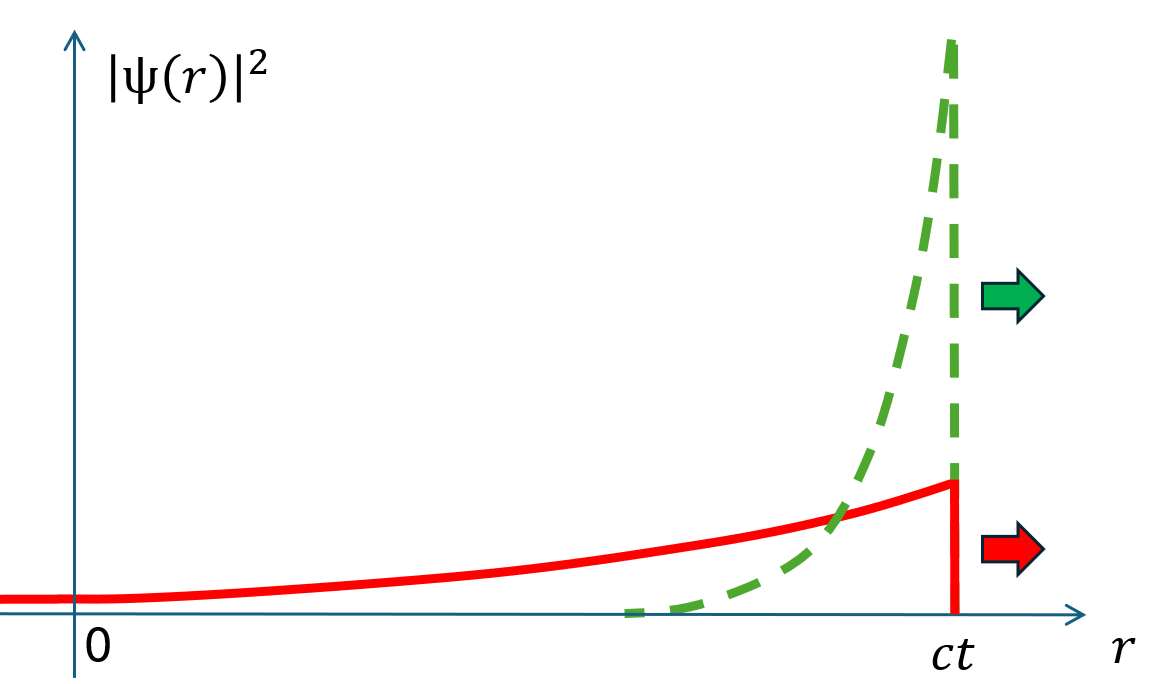} \caption{Squared moduli of non-oscillating neutrino wave functions at time $t$. Both wave packets are emitted at $t=0$ from decaying atoms located at the origin ($r=0$). Their sharp fronts travel at (approximately) the speed of light. Green dashed curve – neutrino originating from a short-lived source. Red solid curve – neutrino produced by a long-lived source, such as (\ref{eq:Cr}). In this case, the wave function near the source has an almost constant amplitude (similar to a plane wave), which gradually decreases with time as $\sim e^{-\Gamma t/2}$. }
\label{fig:Psi}
\end{figure}

\noindent Here, $\mathcal{E}$ denotes the mean neutrino energy, and $\tau = \hbar/\Gamma$ represents the nuclear lifetime. A schematic illustration is given in Fig. \ref{fig:Psi}. Neutrinos originating from rapidly decaying sources (green dashed line) can be located at the origin only within the short time interval $[0,\tau]$. In contrast, neutrinos produced in electron-capture decays of long-lived nuclei (solid red line) may remain in the vicinity of the source for many days. Close to the nucleus, the wave function can be approximated by a plane wave whose amplitude decreases slowly with time. This justifies the use of plane waves in our analysis. 

An oscillating neutrino is a coherent superposition of three mass eigenstates, and the specific internal configuration of these components plays a crucial role. A range of treatments has examined whether the mass eigenstates ought to be assigned equal momentum \cite{Bilenky-1978}, equal energy \cite{Lipkin}, equal velocity \cite{Takeuchi}, or instead be described by a more general kinematic framework \cite{Winter}. The dominant view is that, despite these different choices, the standard oscillation formula remains unchanged \cite{Akhmedov-2009, Cohen}.

In the momentum space, plane waves are described by delta functions. We now select three mass components of our neutrino state in the following way

\begin{align}
\psi_j(p,t)  &= \sqrt{2 \pi} U_{e j}^* \delta(p-\tilde{p}-\chi_j) e^{-i\omega_j t}, \label{eq:25} \\
j &= 1,2,3. \nonumber
\end{align}

\noindent Here $\tilde{p} = \mathcal{E}/c$ denotes the average neutrino momentum. For definiteness, we focus on the decay process given in Eq. (\ref{eq:Cr}), which has a lifetime of 40 days and produces monochromatic neutrinos with an energy of $\mathcal{E} = 0.746$ MeV. In the neighborhood of this value we will employ a linear approximation for the energy-momentum relationship

\begin{align*}
\omega_{j}(p) &= \sqrt{m_j^2+p^2} \approx  p + \delta_{j}, \\
\delta_j &= \frac{m_j^2}{2 \mathcal{E}}, 
\end{align*}

\noindent where $m_j$ denote the neutrino masses. In eq. (\ref{eq:25}), the $\chi_j$ represent small parameters of the same order as $\delta_j$. By varying these parameters, one can readily move between the equal-momentum, equal-energy, and other kinematic regimes. For the coefficients multiplying the delta functions, we have selected the matrix elements $U_{ej}^*$ of the Pontecorvo–Maki–Nakagawa–Sakata matrix \cite{PMNS}. The rationale for this choice will be clarified shortly.

The $\nu_e$ flavor component can be expressed as a linear combination of the functions in (\ref{eq:25}).

\begin{align*}
& \psi_{e}(p,t) =  \sqrt{2 \pi} \sum_{j=1}^3 |U_{ej}|^2 
\delta(p-\tilde{p}-\chi_j) e^{-i\omega_j t}. 
\end{align*}

\noindent We can now apply a Fourier transform to obtain the $\nu_{e}$ wavefunction in the position representation

\begin{align*}
\tilde{\psi}_{e}(r,t) 
&=  \frac{1}{\sqrt{2 \pi}}\int dp e^{ipr} \psi_{\nu_e}(p,t) \\
&= e^{i\tilde{p}(r-t)} \sum_{j=1}^3 |U_{ej}|^2  e^{i\chi_j r} 
 e^{-i(\chi_j+ \delta_j) t}. 
\end{align*}

\noindent The associated probability distribution is

\begin{align}
|\tilde{\psi}_{e}(r,t)|^2  
&=  \sum_{j=1}^3 |U_{ej}|^4 +  2\sum_{j<k} |U_{ej}|^2|U_{ek}|^2 \times \nonumber \\
&\cos((\chi_j - \chi_k)r - (\chi_j - \chi_k)t - \delta_{jk} t),\label{eq:prob3} \\
\delta_{jk} &\equiv \delta_j - \delta_k. \nonumber
\end{align}

\noindent Observe that the specific choice of coefficients in (\ref{eq:25}) ensures that at $t = r = 0$ the detection probability of $\nu_e$ is $|\tilde{\psi}_{e}(0,0)|^2 = 1$. Thus, our model describes a purely electron-flavor neutrino emitted from the origin. We now proceed to study how this model behaves for various selections of the shift parameters $\chi_j$.

\subsection{Equal momentum state}
\label{ss:equal}

By setting $\chi_j = 0$ in (\ref{eq:25}), we obtain a system where all three neutrino mass eigenstates share the same momentum $\tilde{p}$, but possess different energies

\begin{align*}
\omega_{j}(\tilde{p}) &\approx \tilde{p} + \delta_j. 
\end{align*}

\noindent In this case the probability (\ref{eq:prob3}) is independent of position and varies periodically with time

\begin{align*}
& |\tilde{\psi}_{e}(r,t)|^2 \nonumber \\  
&=  \sum_{j=1}^3 |U_{ej}|^4 +  2\sum_{j<k} |U_{ej}|^2|U_{ek}|^2 \cos(\delta_{jk} t).
\end{align*}

\noindent  The second term represents a superposition of cosine functions whose longest oscillation period is  

\begin{align*}
T_{osc} &= \frac{2 \pi \hbar}{\delta_{21}} = 8.3 \times 10^{-5} \ s \, ,
\end{align*}

\noindent which is much shorter than the typical measurement time in neutrino experiments. As a result, during data collection these cosine contributions effectively average to zero, yielding a probability that is independent of distance.

\begin{align*}
& P_{\nu_e \to \nu_{e}}(r,t) =   \sum_{j=1}^3 |U_{ej}|^4.
\end{align*}

\noindent Such behavior is at odds with the results of oscillation experiments. We therefore conclude that real neutrino states cannot have wave functions that share the same momentum.

\subsection{Equal energy state}
\label{ss:equalen}

Let us now set $\chi_j = -\delta_{j}$ in (\ref{eq:25}). With this choice, the three delta functions are centered at distinct momenta $\tilde{p}-\delta_{j}$. However, these states have identical energies

\begin{align*}
\omega_j(\tilde{p}-\delta_{j}) &\approx   \tilde{p}.  
\end{align*}

\noindent  In this situation, the detection probability has the form of a wave frozen in time

\begin{align}
&|\tilde{\psi}_{e}(r,t)|^2  \nonumber \\ 
&= \sum_{j=1}^3 |U_{ej}|^4 +  2\sum_{j<k} |U_{ej}|^2|U_{ek}|^2 \cos(\delta_{kj}r).\label{eq:osc}
\end{align}

\noindent This is the conventional formula employed to analyze neutrino oscillations in experimental settings. It works well in most situations. Nonetheless, for short baselines ($r \ll cT_{osc}$), it yields a unit probability of detecting an electron neutrino

\begin{align}
|\tilde{\psi}_{e}(0,t)|^2 = 1, \label{eq:one}
\end{align}

\noindent and therefore fails to account for the 20\% deficit observed in gallium anomaly experiments.

\subsection{Neutrino emitted in nuclear decay}
\label{ss:equalde}

Strictly speaking, neither equal momentum nor equal energy condition can be true for neutrinos emitted in two-body decays such as (\ref{eq:Cr}) \cite{Winter}. 
Energy conservation instead demands that the momenta $p_{j}$ of the three neutrino mass eigenstates obey

\begin{align*}
M &= \sqrt{N^2 + p_{j}^2} + \sqrt{m_j^2 + p_{j}^2}  
\end{align*}

\noindent where $M$ and $N$ denote the masses of the parent and daughter nuclei, $^{51}\!\mathrm{Cr}$ and $^{51}\!\mathrm{V}$, respectively. Neglecting
 the tiny terms $\propto m_j^4$, we find that the three neutrino components have different momenta

\begin{align*}
p_{j} &\approx \tilde{p}\sqrt{1-\frac{m_j^2(M^2 + N^2)}{2 \tilde{p}^2M^2}} \approx \tilde{p} - \delta_j + \eta_j 
\end{align*}

\noindent and different energies 

\begin{align*}
 \omega_j(p_{j}) \approx p_{j} + \delta_j= \tilde{p}  + \eta_j. 
\end{align*}

\noindent Here

\begin{align*}
\tilde{p} &\equiv \frac{(M^2 - N^2)}{2 M} \approx \mathcal{E}, \nonumber 
\end{align*}

\noindent and parameters

\begin{align*}
\eta_j &\equiv \frac{m_j^2}{2 M}. 
\end{align*}

\noindent express the contributions of nonzero neutrino masses to the nuclear recoil energy.

The corresponding quantum state can be represented by the wave function (\ref{eq:25}) with $\chi_j = -\delta_{j} + \eta_{j}$. The probability for detecting an electron neutrino flavor in the position space is

\begin{align}
& |\tilde{\psi}_{e}(r,t)|^2 =  \sum_{j=1}^3 |U_{ej}|^4 \nonumber \\
&+  2\sum_{j<k} |U_{ej}|^2|U_{ek}|^2 \cos((-\delta_{jk} + \eta_{jk})r - \eta_{jk}t), \label{eq:eta21a} \\
&\eta_{jk} \equiv \eta_j - \eta_k. \nonumber
\end{align}

\noindent  In contrast to the standard stationary expression (\ref{eq:osc}), this solution exhibits variation in both space and time.  
Specifically, in the vicinity of the source ($r=0$) the oscillatory probability becomes

\begin{align}
&|\tilde{\psi}_{e}(0,t)|^2 \nonumber \\
&=  \sum_{j=1}^3 |U_{ej}|^4 +  2\sum_{j<k} |U_{ej}|^2|U_{ek}|^2 \cos(\eta_{jk}t). \label{eq:eta21y}
\end{align}

\noindent It coincides with the usual result (\ref{eq:one}) only under the assumption that all parameters $\eta_j$ vanish, e.g., in the limit of infinite nuclear mass $M \to \infty$. 

\section{Application to short baseline anomalies}
\label{sc:Application}

\subsection{Gallium anomaly}

Note that for short wave packets produced by a rapidly decaying nuclei and propagating at the speed of light (green dashed line in Fig. \ref{fig:Psi}), the standard oscillation formula (\ref{eq:osc}) can be recovered from eq. (\ref{eq:eta21a}) by performing the “time-to-space conversion” $t = r$ \cite{Winter}. However, this conversion becomes doubtful for electron capture decays such as (\ref{eq:Cr}). In such cases, the source lifetime is on the order of weeks, the probability of finding the neutrino near the nucleus decreases very slowly, and the substitution $t \to r$ is not warranted.

To employ our formula (\ref{eq:eta21y}) for the gallium anomaly experiments, we must account for the fact that actual neutrino measurements are performed over extended periods of time. In the BEST experiment \cite{BEST}, for instance, each data-taking cycle lasted about $t_2 - t_1 \approx 10$ days. In addition, in the vicinity of the source ($r=0$), the probability of detecting a neutrino was modulated by the decay factor $\Gamma e^{-\Gamma t}$. Consequently, the overall probability has to be determined by integrating over the observation interval $[t_1,t_2]$

\begin{align}
&P_{\nu_e \to \nu_{e}}^{recoil}(t) = \Gamma \int _{t_1}^{t_2} e^{-\Gamma t}|\tilde{\psi}_{e}(0,t)|^2 dt \nonumber  \\ 
&= \Gamma  \int _{t_1}^{t_2} e^{-\Gamma t} \times \nonumber \\
& \left[\sum_{j=1}^3 |U_{ej}|^4 +  2\sum_{j<k} |U_{ej}|^2|U_{ek}|^2 \cos(\eta_{jk}t) \right]  dt.\label{eq:eta22} 
\end{align}

\noindent Given that the longest oscillation period is $T_{recoil} = 2 \pi \hbar/\eta_{21} = 5.3\,\text{s}$, the observation interval is $[t_2 - t_1] = 10$ days, and the source lifetime is $\tau = 40$ days, we have

\begin{align*}
T_{recoil} \ll [t_2 - t_1] < \tau \, .
\end{align*}

\noindent Under these conditions, in eq. (\ref{eq:eta22}) the average factor $e^{-\Gamma \bar{t}}$ (with $\bar{t} \equiv (t_2 + t_1)/2$) can be taken outside the integral, and the rapidly oscillating cosine terms may be neglected

\begin{align}
& P_{\nu_e \to \nu_{e}}^{recoil}(\bar{t}) \approx \Gamma  e^{-\Gamma \bar{t}} [t_2-t_1] \sum_{j=1}^3 |U_{ej}|^4.  \label{eq:integrand2}
\end{align}

In the standard treatment of oscillations, where nuclear recoil is ignored ($M = \infty, \eta_{jk}=0$), the square bracket in (\ref{eq:eta22}) becomes unity, and

\begin{align}
P_{\nu_e \to \nu_{e}}^{no \ recoil}(0,\bar{t})&= \Gamma e^{-\Gamma \bar{t}} [t_2-t_1]. 
\label{eq:their}
\end{align}

\noindent When this is applied to the gallium anomaly experiments, both (\ref{eq:integrand2}) and (\ref{eq:their}) correctly reproduce the exponential decrease of the detection rate \cite{SAGE, BEST-2022, Gallex-1998}. Our expression (\ref{eq:integrand2}), however, additionally predicts a deficit of electron neutrinos in the vicinity of the source

\begin{align*}
\varepsilon \equiv 1- \frac{P_{\nu_e \to \nu_{e}}^{recoil}(0,\bar{t})}{P_{\nu_e \to \nu_{e}}^{no \ recoil}(0,\bar{t})} = 1- \sum_{j=1}^3 |U_{ej}|^4 = 0.45. 
\end{align*}

\noindent See Table \ref{table:2.2x}. This result exceeds the measured value of $20\%$, suggesting that nuclear recoil by itself cannot fully account for the observation.

\begin{table*}[ht]
\caption{Short baseline (anti)neutrino anomalies.}
\begin{tabular}{|c|ccc|}
\hline
     & gallium ($^{51}\!\mathrm{Cr}$)   & \hspace{0.2 in} reactor \hspace{0.2 in} & LSND ($\mu^{+}$) \hspace{0.2 in}\\
\hline
$\tau = \hbar/\Gamma$ &   40 days & 1-10 s & $2.2 \times 10^{-6}$ s \\
$2 \pi \hbar/\eta_{21}$, s & 5.3  &  $\sim 10$ & 0.012  \\
$2 \pi \hbar/\eta_{31} \approx 2 \pi \hbar/\eta_{32} $, s & 0.16  &  $\sim 0.3$ & $3.5 \times 10^{-4}$  \\
Anomaly $\varepsilon$, theor., \% & 45  &  15-44 & 0.0035 \\
Anomaly $\varepsilon$, exp., \% & 20 \cite{Gallium_anomaly} &  4.5 \cite{Giunti-2026}& 0.26 \cite{LSND}  \\
\hline
\end{tabular}
\label{table:2.2x}
\end{table*} 

The nuclear recoil contribution to the short-baseline neutrino deficit is anticipated to be similar when using other long-lived sources, such as $^{65}\!\mathrm{Zn}$ \cite{Gallium_anomaly} and $^{58}\!\mathrm{Co}$ \cite{Gavrin-2025}, or when employing different detector technologies \cite{Bellini-2013, Huber, Benato, Chauhan, Ciuffoli, Semenov}. Therefore, these proposed experiments could help clarify which additional physical mechanisms play a role in the gallium anomaly.

\subsection{Reactor antineutrino anomaly}

Experiments using reactor antineutrinos have revealed a short-baseline flux deficit of 4.5\% \cite{Giunti-2026}. These measurements differ in several important respects from the gallium anomaly experiments. Reactor antineutrinos originate from three-body beta decays and therefore do not have the monochromatic energy spectrum characteristic of the two-body reactions like (\ref{eq:Cr}).  
The interpretation of reactor data is further complicated by the presence of hundreds of contributing isotopes, whose lifetimes span from fractions of a second to millions of years and which produce antineutrinos in a broad energy range. In principle, all such contributions should be combined with appropriate weights. In this paper, however, we restrict ourselves to an order-of-magnitude estimate.  

Most antineutrinos with energies above the 1.8 MeV detection threshold come from decays of early stage fission fragments with lifetimes in the range 1–10 s. These fragments typically have masses clustered around $M = 100 \ \text{a.m.u.}$, and their recoil oscillation periods $2 \pi \hbar/ \eta_{jk}$ are of the same order as their lifetimes. See Table \ref{table:2.2x}.

In reactor-based experiments, one works with a steady flux of neutrinos. Then the observation period is much longer than other characteristic time scales

\begin{align*}
T_{recoil} \approx \tau \ll [t_2-t_1],
\end{align*}

\noindent and the integration limit in (\ref{eq:eta22}) can be extended to $[0, \infty]$

\begin{align*}
 P^{recoil}_{\tilde{\nu}_e \to \tilde{\nu}_{e}} 
&=  \sum_{j=1}^3 |U_{ej}|^4 +  2\sum_{j<k} |U_{ej}|^2|U_{ek}|^2  \frac{\Gamma^2}{\Gamma^2 + \eta_{jk}^2}.
\end{align*}

\noindent   For lifetimes $\tau$ in the interval from $1\,\text{s}$ to $10\,\text{s}$, the calculated anomaly

\begin{align*}
\varepsilon &= 1 - P^{recoil}_{\tilde{\nu}_e \to \tilde{\nu}_{e}} \label{eq:zeta}
\end{align*}

\noindent lies between 15\% and 44\%, which overestimates the experimental value. See Table~\ref{table:2.2x}.

\subsection{LSND anomaly}

In the LSND experiment \cite{LSND}, muon antineutrinos were produced from muons decaying at rest,

\begin{align*}
\mu^+ \to e^+ + \nu_e + \tilde{\nu}_{\mu}. 
\end{align*}

\noindent In this setup, an unexpectedly large number of electron antineutrinos was observed over a short baseline of 30 meters.
To study the $\tilde{\nu}_{\mu} \to \tilde{\nu}_{e}$ transition probability, we begin with an initial $\tilde{\nu}_{\mu}$ state expressed in terms of its mass eigenstate components as follows

\begin{align*}
\psi_j(p,t)  &= \sqrt{2 \pi} U_{\mu j}^* \delta(p-\tilde{p}+\delta_j - \eta_j) e^{-i\omega_j t}. 
\end{align*}

\noindent We are interested in the $\tilde{\nu}_e$ flavor wave function

\begin{align*}
\psi_{e}(p,t)  &= \sqrt{2 \pi} \sum_{j=1}^3 U_{e j}U_{\mu j}^* \delta(p-\tilde{p}+\delta_j - \eta_j) e^{-i\omega_j t}. 
\end{align*}

\noindent The value of the position space wave function at the origin is 

\begin{align*}
\tilde{\psi}_{e}(r=0,t)  &= e^{-i \tilde{p}ct }\sum_{j=1}^3 U_{e j}U_{\mu j}^* e^{-i\eta_j t}. 
\end{align*}

\noindent Then, assuming that matrix elements of the PMNS matrix are real, the probability of observing a $\tilde{\nu}_e$ neutrino at short distances is given by

\begin{align*}
P^{recoil}_{\tilde{\nu}_{\mu} \to \tilde{\nu}_{e}} 
&= \sum_{j=1}^3 U_{e j}^2 U_{\mu j}^2 + 2 \sum_{j<k} U_{e j} U_{\mu j} U_{e k} U_{\mu k} \frac{\Gamma^2}{\Gamma^2 + \eta_{jk}^2} \\
&= 3.5 \times 10^{-5} \;.
\end{align*}

\noindent This value is two orders of magnitude smaller than the experimental result reported in \cite{LSND}. See Table~\ref{table:2.2x}. Consequently, the recoil effect provides only an insignificant contribution to the LSND anomaly.

\section{Conclusions}
\label{sec:discussion}

Nuclear recoil affects the distribution of momentum and energy among neutrino components produced in decays. Consequently, the probability of detecting these neutrinos exhibits periodic modulations with characteristic periods ranging from $T_{recoil}\sim 3.5 \times 10^{-4}$ s for $\tilde{\nu}_{\mu}$ in the LSND experiment to $T_{recoil}\sim 10$ s for $\tilde{\nu}_e$ emitted by reactor isotopes. Thus, the detection probability at a fixed spatial point is determined not only by the standard parameter $r/\mathcal{E}$, but also by how $T_{recoil}$ relates to the experimental observation window $[t_2 - t_1]$ and to the source lifetime $\tau$.

In this paper, we examined how recoil-induced oscillations influence neutrino measurements at short baselines near stationary sources. Our analysis indicates that this effect is negligible for short-lived (anti)neutrino sources, such as muons in the LSND experiment. 
At the same time, our calculations overestimate the sizes of the gallium and reactor anomalies, indicating that a complete explanation of these effects is still lacking. Nonetheless, it is clear that nuclear recoil has to be included in any discussion of these important phenomena.

\end{document}